\documentclass[10pt,superscriptaddress,showkeys]{revtex4}
\newcommand{\eq}[1]{Eq.~{(\ref{#1})}}
\newcommand{\fig}[1]{Fig.~{\ref{#1}}}
\newcommand{\bea}{\begin{eqnarray}}
\newcommand{\beann}{\begin{eqnarray*}}
\newcommand{\eea}{\end{eqnarray}}
\newcommand{\eeann}{\end{eqnarray*}}
\newcommand{\SZ}{Sunyaev-Zel'dovich\ }

\usepackage{graphicx}
\usepackage{amssymb}
\begin{document}
\title{Impact of Systematic Errors in \SZ Surveys of Galaxy Clusters}
\author{Matthew R. Francis}
\email[E-mail: ]{mfrancis@physics.rutgers.edu} \affiliation{Dept. of
  Physics and Astronomy, Rutgers University, 136 Frelinghuysen Road,
  Piscataway, NJ 08854}
\author {Rachel Bean}\email[E-mail: ]{rbean@astro.cornell.edu}
\affiliation{Department of Astronomy, Space Sciences Building, Cornell
  University,  Ithaca, NY 14853}
\author{Arthur Kosowsky} \email[E-mail: ]{kosowsky@pitt.edu}
\affiliation{Dept. of Physics and Astronomy, University of Pittsburgh,
3941 O'Hara Street, Pittsburgh, PA 15260}
\date{\today}

%%%%%%%%%%%%%%%%%%%%%%%%%%%%%%%%%%%%%%%%%%
\begin{abstract}
%%%%%%%%%%%%%%%%%%%%%%%%%%%%%%%%%%%%%%%%%%
Future high-resolution microwave background measurements hold the
promise of detecting galaxy clusters throughout our Hubble volume
through their \SZ (SZ) signature, down to a given limiting flux. The number
density of galaxy clusters is highly sensitive to cluster mass through
fluctuations in the matter power spectrum, as well as redshift through
the comoving volume and the growth factor. This sensitivity in
principle allows tight constraints on such quantities as the equation
of state of dark energy and the neutrino mass.  We evaluate the
ability of future cluster surveys to measure  these quantities
when combined with Planck-like CMB data. Using
a simple effective model for uncertainties in the cluster mass-SZ flux
relation, we evaluate systematic shifts in cosmological constraints
from cluster SZ surveys. We find that a systematic bias of 10\% in
cluster mass measurements can give rise to shifts in cosmological
parameter estimates at levels larger than the $1\sigma$ statistical
errors. Systematic errors are unlikely to be detected from the mass
and redshift dependence of cluster number counts alone; increasing
survey size has only a marginal effect. Implications
for upcoming experiments are discussed.
\end{abstract}
\maketitle

%%%%%%%%%%%%%%%%%%%%%%%%%%%%%%%%%%%%%%%%%%
\section{Introduction}
%%%%%%%%%%%%%%%%%%%%%%%%%%%%%%%%%%%%%%%%%%

Galaxy cluster surveys have the potential to place strong constraints
on cosmological models, as has long been appreciated 
\cite{white93,viana96,eke96,barbosa96,bahcall97,bahcall98,mei04,molnar04}.  
The number of galaxy clusters as a function of cluster
mass and redshift depends sensitively on both the gravitational growth
factor and the comoving volume element as the universe evolves
\cite{colafrancesco97,holder01,kneissl01,benson02,diego02,weller02}.
In principle, large catalogs of thousands of galaxy clusters with masses
and redshifts will constrain all cosmological parameters which affect
either the growth factor or the rate of expansion at redshifts below
$z\simeq 2$, where significant numbers of clusters have formed. In
particular, this includes $w$, representing the equation of state of
the dark energy, and $\sum m_\nu$, the total mass of the three neutrino
species. Both of these quantities are of crucial importance for
fundamental physics, and neither is well constrained by measurements
of the cosmic microwave background power spectrum. Clusters will also
give constraints on the total mass density of the universe,
$\Omega_m$, and the Hubble parameter $h$ which are complementary to
those from the microwave background.

In practice, the difficulty with using clusters as a cosmological
probe lies in estimating their masses and in obtaining complete
cluster samples. Past cluster surveys have relied on either X-ray or
optical selection, but such surveys are complete only in a comparatively
local region of the universe, and the connection between the observed
optical richness or X-ray luminosity and the mass of the cluster is
difficult to determine. With detection of clusters via their thermal
\SZ (SZ) distortion of the microwave background now
firmly established (see \cite{carlstrom02} for a fairly recent
comprehensive list of detections), large cluster surveys with
excellent completeness and improved selection functions are imminent
\cite{act03,act04,spt04,amiba01,ami02}.  (For reviews of the SZ
effect, see \cite{sz1969,sz1970,sunyaev80,rephaeli95,
  birkinshaw99,carlstrom02}).  Since the SZ effect is essentially
independent of redshift, SZ surveys will provide all galaxy clusters
in the direction of a given sky region down to a limiting SZ
distortion depending on cluster gas mass and pressure. Furthermore,
the connection between SZ distortion and cluster mass is likely less
sensitive to internal cluster physics than  the corresponding
connection for X-ray luminosity \cite{motlSZ2005}. Experimental
advances in detecting
microwave fluctuations at small angular scales have raised hopes that
we will soon have cluster catalogs in hand which will give meaningful
information about fundamental physics.

While a number of papers so far have estimated the statistical errors in
cosmological parameters from cluster surveys, relatively less emphasis 
has been placed on systematic errors and their 
impact (although see \cite{hmh,holder01,molnar04,huterer04}). An early
important contribution \cite{holder01} concluded that
realistic uncertainties in cluster masses can lead to significant
biases in cosmological parameters. Several subsequent papers, however,
have implied that systematic errors may not be a limiting factor
for cosmological conclusions 
\cite{wkhm2004,battye_weller,melin2005}. It is clear that our ability to 
construct an unbiased estimator of cluster masses, whether through a
correlation with their SZ signal, from gravitational lensing, or
through other observables, will be the most important factor in
determining the cosmological utility of cluster SZ
catalogs. Conversely, as large SZ cluster surveys are now being 
planned, it is important to have a target accuracy for cluster mass
determination: this affects not only observation strategies for the SZ
signal, but also for the kinds of follow-up observations in other wave
bands which will be required.

Our intention in this paper is to build on previous work
\cite{holder01}  and focus on the effects of systematic errors on
cosmological parameter extraction in the context of statistical
constraints from upcoming microwave experiments. This paper aims
to give a quantitative analysis of how well we need to understand
cluster properties in order to realize the cosmological 
potential of these SZ cluster surveys. In particular, how small must
systematic errors in cluster mass and redshift estimates be so that,
for a cluster catalog of a given size, the bias in cosmological
parameter determination due to the systematic errors is smaller than
the statistical errors?  We tackle this issue without relying upon any
particular assumptions about the detailed cluster physics and
survey-specific issues like cluster selection functions.  Instead, we directly
consider the uncertainties in the cosmologically relevant quantities,
namely the cluster mass and redshift. This paramaterization
effectively encompasses any uncertainty in cluster physics. 

The following Section reviews the formalism for generating mock
cluster catalogs from a given cosmological model, and displays
the cluster distribution in both mass and redshift for several
underlying cosmologies. Section~\ref{clust_systematics} discusses some
details of model fitting and error determination, using Monte Carlo
and parameter-space search techniques. Then in
Section~\ref{clust_results} we present results for the error in
cosmological parameters due to biased cluster mass determination, for
a number of different bias levels. We also display cluster mass and
redshift distribution residuals between the best-fit model and the
mock cluster catalog, for different assumptions about the cluster mass
error. With large enough observed cluster catalogs, small differences
in the observed distribution and the cluster distribution from the
best-fit cosmological model can be statistically significant; we
quantify the size of samples needed to detect these discrepancies. The
concluding Section discusses these results in context of upcoming
microwave cluster surveys.  

%%%%%%%%%%%%%%%%%%%%%%%%%%%%%%%%%%%%%%%%%%
\section{Dependence of Cluster Evolution on Cosmological Parameters}
\label{clust_evolution}
%%%%%%%%%%%%%%%%%%%%%%%%%%%%%%%%%%%%%%%%%%

Ideally, a cluster \SZ survey will identify all the clusters in a
certain angular region of the sky, $\delta\Omega$, and find their
masses, $M$, and redshifts, $z$.  The method of estimating the cluster
distribution is well known (see, \emph{e.g}, \cite{hmh}).  Consider
the comoving mass function, which is the number density of clusters 
\bea
\frac{dN}{dM dz} (M, z) = \delta\Omega \frac{dV}{dz d\Omega} (z)
\frac{dn}{dM} (M,z)
\label{comassfunc}
\eea
within the comoving volume element $dV/dzd\Omega$ for a given solid
angle $\delta\Omega$ on the sky.  The mass function
\bea
\frac{dn}{dM} (M, z) &=& 0.315 \frac{\rho_0}{M^2} \frac{d \ln
  \sigma_M}{d \ln M}  \times \exp \left\{ - \left| 0.61 - \ln
(\sigma_M D_z) \right|^{3.8}\right\} 
\label{jenkins_mf}
\eea
describes the number density, $n$, of objects between masses of $M$ and $M+
dM$ at a given redshift $z$, where $\rho_0$ is the present density of
matter.  \eq{jenkins_mf} is obtained from $N$-body cluster simulations
\citep{jenkins2001} assuming a standard cosmological model. The
dependence on mass comes through the spherical over-density 
\bea
\sigma_M{}^2 = \int^\infty_0 dk (4\pi k^2) P(k) W^2(k R(M)),
\eea
where the matter power spectrum $P(k)$ is integrated within a
sharply-defined spherical region of radius $R$, containing mass $M = 4
\pi \rho_0 R(M)^3/3$ with a top-hat window function $W(x)$.  The
mass-independent quantities in Eqs.~(\ref{comassfunc})
and~(\ref{jenkins_mf}) are the volume factor $dV/dzd\Omega$ and the
linear growth function $D_z = \delta(z)/\delta(0)$, where \[\delta(z)
= H(z) \int^{(1+z)^{-1}}_0 \frac{da}{(a H(a))^3} .\]  For a given
cosmology, the above equations completely determine the cluster
abundance. Note that small changes in the mass fluctuations
$\sigma_M$, specifically slight variations in numerics or in how the
window function is defined, can lead to significant variations in
$dn/dM$ due to its exponential dependence on $\sigma_M$.  

Neutrinos and dark energy have complementary effects, based on how
they enter into \eq{comassfunc}. Dark energy has little effect on the
primordial power spectrum, but directly affects the volume and growth
factors.  Neutrinos leave the volume factor unchanged, but suppress growth
of fluctuations on scales smaller than the neutrino free-streaming length; total neutrino masses on the order of 0.5 eV can
substantially suppress the power spectrum at scales relevant to
cluster physics (\emph{e.g.}, $k \gtrsim 0.02\ h/\mathrm{Mpc}^{-1}$).  

\begin{figure}
\begin{center}
  \includegraphics[width=3.3in]{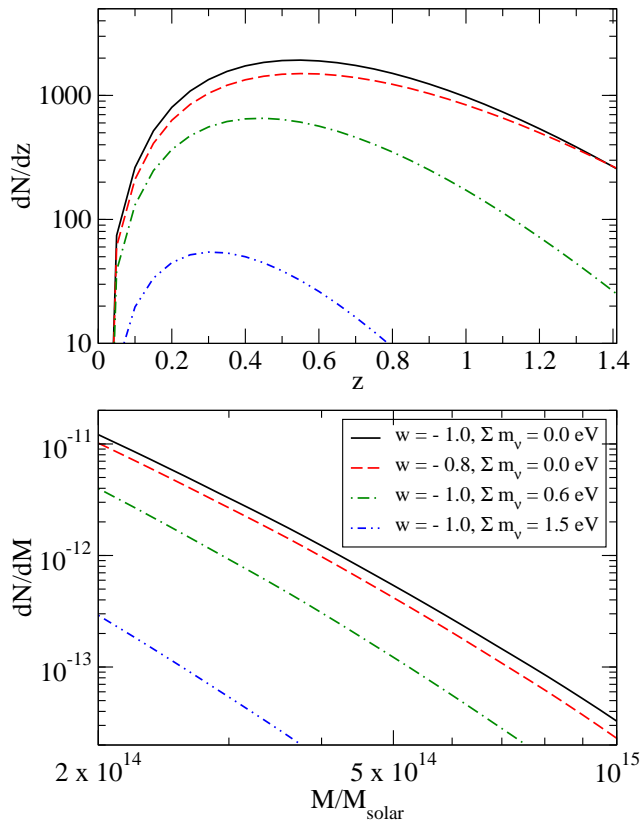}
\end{center}
  \caption[Cluster mass function integrated over $M$ and $z$]{The mass
  function integrated over mass (from $M_{lim} = 2 \times 10^{14}
  M_\odot$, top) and over redshift (bottom), for a survey area of $\delta\Omega = 200$
  square degrees.  Note that even a relatively small neutrino mass
  suppresses cluster formation strongly.} 
  \label{massfunc}
\end{figure}

Frequently, the cluster density is integrated over mass to yield the
total cluster density in redshift only: 
\bea
\frac{dN}{dz} (z; M_{lim}) = \int^\infty_{M_{lim}} dM \frac{dN}{dM dz},
\label{mass_cutoff}
\eea
where the lower limit is an experimentally-determined limiting mass
(which generally should depend on redshift \citep{hmh}).  This is the
quantity plotted in the first plot of \fig{massfunc}.  However, a real
survey will contain information about cluster masses through a
flux-mass relation (see \emph{e.g.} \cite{bryan_norman}), so
neglecting the mass dependence loses information that can potentially
be used to constrain cosmological parameters.  Therefore, we also
consider binning using the distribution function given in
\eq{comassfunc}: 
\bea
N_{ij} = \int_{M_i} dM \int_{z_j} dz \frac{dN}{dM dz} 
\label{binning}.
\eea

A given galaxy cluster survey will provide an estimate of $N_{ij}$;
the remainder of this paper considers the impact of systematic mass
errors in this estimate on the cosmological conclusions which can be
drawn from it. 

%%%%%%%%%%%%%%%%%%%%%%%%%%%%%%%%%%%%%%%%%%
\section{Treatment of Systematic Error}
\label{clust_systematics}
%%%%%%%%%%%%%%%%%%%%%%%%%%%%%%%%%%%%%%%%%%

The previous section describes how, given a set of cosmological
parameters, it is possible to obtain the theoretical distribution of
clusters in mass and redshift.  For real cluster catalogues, it is
necessary to consider the converse procedure, taking a set of measured
cluster counts as in \eq{binning} and constraining cosmological
parameters from it.  Markov Chain Monte Carlo techniques are well
established in cosmology for constraining multi-dimensional parameter
spaces \cite{christensen2001,kmj,lewis_bridle,wmapy1_method}.
The cosmological information in an SZ survey depends on the minimum
cluster mass probed by the survey and the survey's angular
coverage.  The following analysis considers a Planck-like measurement of the microwave
background primordial power spectrum combined with an ACT-like \SZ
cluster survey.

The utility of a given survey, however, depends on the extent to which systematic errors
affect the inferred parameter values for a particular cluster catalogue.
Cluster masses and redshifts are the fundamental quantities which can be determined
for a given cosmological model from numerical simulations, and we will extract cluster mass
and redshift estimates from upcoming observations.
While redshifts can be measured to high accuracy with sufficient telescope
time, determining masses poses a significantly harder challenge.  Poorly
understood galaxy cluster physics which modify a cluster's SZ signature
can be viewed as a potential systematic
error in mass estimates based on the SZ signal itself.  Quantifying the effect of such systematic
errors on constraining cosmological parameters is the main goal of
this analysis.

Much of the cluster gas physics which is difficult to model---shock heating of
intracluster gas, feedback from supernovae and active galactic nuclei
\cite{AGN_heating}, magnetic field turbulence
\cite{dolag,mccarthy03a}---has the tendency to increase the SZ flux
of a cluster relative to its mass, an effect observed especially in
low-mass clusters (the ``entropy floor''
\cite{ponman99,lloyd00,mccarthy03b}). Therefore, the effect of these
systematics 
%is to make clusters in \SZ surveys appear more massive than they are,
is to make cluster masses inferred from their \SZ distortion larger
than the actual cluster mass,
boosting clusters from lower-mass bins into higher-mass bins, and
increasing the total number of clusters in the sample above 
what would na\"{\i}vely be expected from the sharp mass cutoff of
\eq{mass_cutoff}.  (Cooling flows, which we do not consider here, have
the opposite effect, decreasing the SZ signature in relation to mass
\cite{cooling_flows,reeseSZ2002}.)  A simple first-order model for the
measured mass, motivated by numerical simulations \cite{dasilva}, is 
\bea
M = M_{real} (1 + s),
\label{mass_shift}
\eea
where $s \geq 0$ is a constant. We neglect any statistical errors in
the mass estimate, which in practice expand the error region in
parameter space without moving its central value; such errors will
also tend to increase the number of clusters in high-mass bins.  (Lima
and Hu \cite{lima_hu2005} consider a scatter in the flux-mass
relationship, which also produces significant effects.)

\begin{figure*}
\begin{center}
  \includegraphics[width=7in]{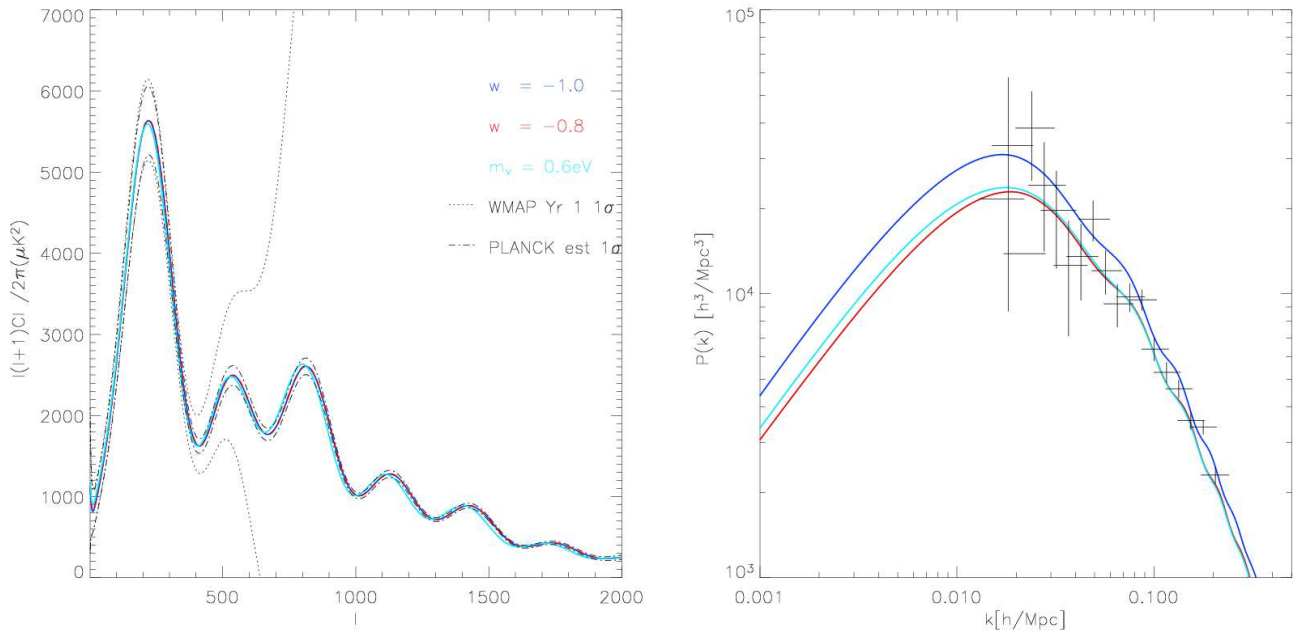}
\end{center}
  \caption{Fiducial CMB and matter power spectra for models A,B and C
    as described in the text. Also shown are the error bars from the
    1st year of WMAP data \cite{wmapy1}, projected errors for the
    Planck experiment, and data from the SDSS galaxy survey
    \cite{Tegmark04}}
  \label{fidspec}
\end{figure*}

We calculate the CMB and matter power spectra
using the CAMB \cite{camb} code for a cosmological scenario involving
7 parameters: $\Omega_bh^2,\Omega_{CDM}h^2, h, \tau,n_s,A_s$, and either
$w$ (assumed not to evolve with redshift) or $\sum m_\nu$.  In
addition, $\Omega_{total} = 1$ is held fixed.  From a set of fiducial
parameters, we calculate the temperature $C_{l}$ and cluster
$N_{ij}$ assuming a Planck-style CMB experiment and an ACT-style SZ
cluster survey over 200 square degrees out to $z=1.4$ with actual
limiting mass $2\times 10^{14}M_{\odot}$ and Poisson error
bars. Systematic mass error is
introduced by taking the ideal cluster binned data and relabeling the
mass bins according to the prescription
$M_{real} = M/(1+s)$, so that the total number of clusters in the
survey is invariant.  Then we determine which cosmological models are consistent with this
altered cluster distribution by a standard Markov chain Monte Carlo
calculation.  Using the prescription given in
\cite{variance_limited} for CMB likelihoods and Poisson error bars for
the cluster counts, we obtain the $\chi^2$ between the best-fit model
and the altered cluster distribution.  Statistical errors scale with the square root
of the number of clusters $\sqrt{N_{ij}}$, or equivalently the survey
area $\sqrt{\delta\Omega}$.

We actually use the total number of clusters in each redshift bin,
Eq.~(\ref{mass_cutoff}), as our observable for the cluster likelihood,
rather than breaking the distribution into a number of mass bins as
described by \eq{binning}. This is because the dominant constraint on the
parameters comes from the Planck-like CMB spectrum, and we find that
breaking the cluster data into a number of mass bins does not
significantly alter the error bars. The mass-binned distribution
$N_{ij}$ is still potentially useful for
assessing goodness-of-fit for a given model, as discussed below.
Real data would have a scatter in the number of clusters in each
redshift bin  consistent with Poisson errors; this scatter
is neglected here so that the effect of any systematics on parameter
determination is isolated from statistical error.  
%If the best-fit
%$\chi^2$ is sufficiently large, the model fails to fit well, and the
%presence of systematics may be detectable over Poisson error, but for
%small $\chi^2$, the systematics are hidden and must be ferreted out by
%some other method.  This issue will be discussed further in
%Section~\ref{clust_results}.

%%%%%%%%%%%%%%%%%%%%%%%%%%%%%%%%%%%%%%%%%%
\section{Numerical Results}
\label{clust_results}
%%%%%%%%%%%%%%%%%%%%%%%%%%%%%%%%%%%%%%%%%%

\begin{figure}
\begin{center}
  \includegraphics[width=7in]{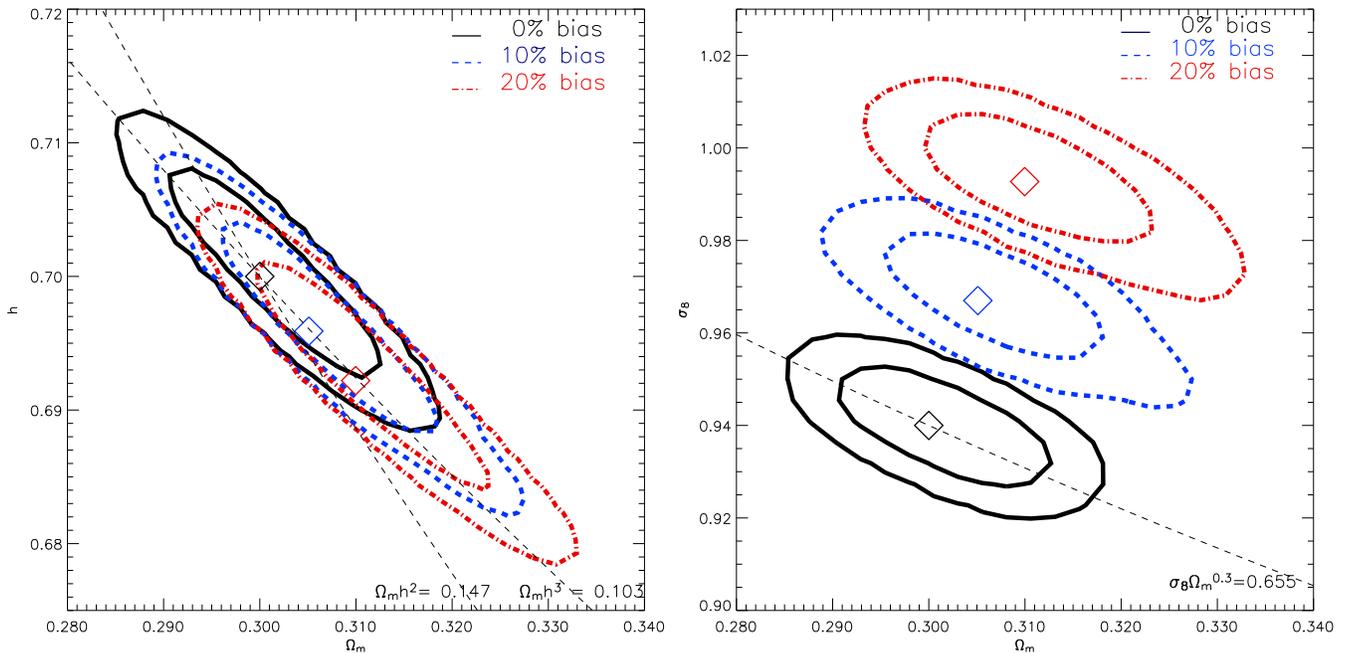}
\end{center}
  \caption{The effect on key cosmological parameters of 0\%, 10\% and
    20\% positive bias in the cluster mass estimate for the $\Lambda$CDM model.} 
  \label{lcdm1}
\end{figure}

\begin{figure*}
\begin{center}
  \includegraphics[width=4.5in]{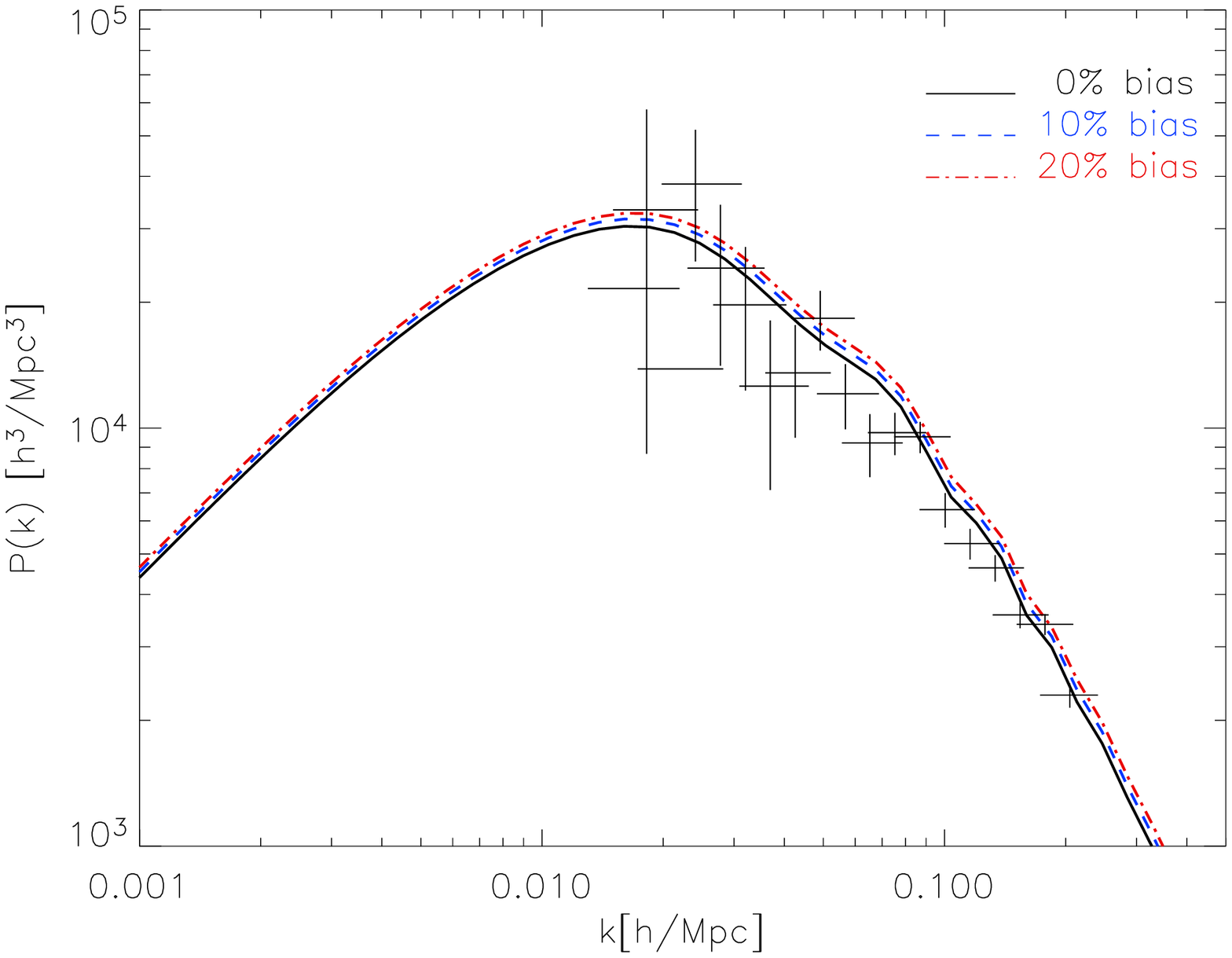}
\end{center}
  \caption{The effect on the best fit matter power spectrum P(k) of
    imposing 0\%,10\% and 20\% cluster mass misestimation in the
    $\Lambda$CDM model. As one would expect the effect of the bias is
    analogous to a boost in power on cluster scales.}
  \label{lcdm2}
\end{figure*}

The fiducial cosmological parameters are chosen to give CMB spectra
closely degenerate with the best fit spectrum from WMAP \cite{wmapy1},
consistent with projected error bars from Planck,  and with the matter
power spectrum from SDSS \cite{Tegmark04}, as shown in \fig{fidspec}.
The effect of adding a bias to a fiducial $\Lambda$CDM scenario is
shown in \fig{lcdm1}. A positive bias in the cluster mass estimate
corresponds to believing that  massive clusters are more
numerous than is actually the case. The
bias mimics an increased growth on cluster scales and
roughly translates into an inferred increase in the density of
clustering matter, increase in spectral tilt, or increase in overall
amplitude. Including Planck-like CMB temperature data tightly
constrains the tilt along with $\Omega_{b}h^{2}$ and $\Omega_{c}h^{2}$, with
the latter implying that the bias might also drive an associated
reduction in the inferred Hubble constant, $H_{0}$. The quantitative
shift in the best fit cosmological parameters produced by systematic
mass errors ($s = 0, 0.1, 0.2$) is shown in table~\ref{param_table}. We
find a shift on the order of 1$\sigma$ in all parameters except for
$\sigma_{8}$, which moves roughly 6$\sigma$. However, the $\sigma_{8}$
values are all within the 1$\sigma$ region
obtained from the combined WMAP year 1 + SDSS matter power spectrum
data ($\sigma_{8}= 0.917^{+0.090}_{-0.072}$)\cite{Tegmark04}; this
implies that the effect of the bias would not be significantly better
constrained by including current matter power spectrum data. The
best-fit linear $P(k)$ for the scenarios with and without bias are shown
in \fig{lcdm2}.

\begin{table*}[t]
\begin{center}
  \begin{tabular}{|c|c|c|c|c|c|c|c|c|}
  \hline
    & \rule[-3mm]{0mm}{8mm} & \multicolumn{7}{|c|}{\textbf{Best-Fit
    Parameter}} \\[0.5ex] \cline{3-9}
    \raisebox{3mm}[0pt]{\textbf{Model}} &
    \raisebox{3mm}[0pt]{\ \textbf{Error}\ } & $h$\rule[-3mm]{0mm}{8mm} &
    $\Omega_b$ & $\Omega_{CDM}$ & $\sum m_\nu$ (eV) & $\Omega_m$ & $w$
    & $\sigma_8$ \\[0.5ex] 
     \hline
% Model LCDM (no neutrinos, w = -1)
    \rule[-3mm]{0mm}{8mm} & 0\% &\ 0.70 $^{+ 0.01}_{-0.01}$ \ &\ 0.050$^{+ 0.001}_{-0.001}$  \ &\ 0.25$^{+ 0.01}_{-0.01}$  \ & &\
    0.30$^{+ 0.01}_{-0.01}$  \ &\  \ &\ 0.93$^{+ 0.01}_{-0.01}$  \  \\ \cline{2-5}\cline{7-7}\cline{9-9} 
    \raisebox{3mm}[0pt]{Model $\Lambda$CDM}\rule[-3mm]{0mm}{8mm} & 10\% & 0.69 &
    0.050 & 0.25 & 0.0 & 0.30 & -1.00 & 0.97  \\
    \cline{2-5}\cline{7-7}\cline{9-9} 
    \raisebox{3mm}[0pt]{\ $\sim$1942 clusters} \rule[-3mm]{0mm}{8mm} &
    20\% & 0.69 & 0.051 & 0.26 & & 0.31 & & 0.99 \\
    \hline\hline
% Model A (no neutrinos, w = -1)
    \rule[-3mm]{0mm}{8mm} & 0\% &\ 0.70 $^{+ 0.02}_{-0.03}$ \ &\ 0.050$^{+ 0.005}_{-0.002}$  \ &\ 0.25$^{+ 0.01}_{-0.03}$  \ & &\
    0.30$^{+ 0.01}_{-0.03}$  \ &\ -1.00$^{+ 0.05}_{-0.04}$  \ &\ 0.93$^{+ 0.03}_{-0.03}$  \  \\ \cline{2-5}\cline{7-9} 
    \raisebox{3mm}[0pt]{Model A}\rule[-3mm]{0mm}{8mm} & 10\% & 0.67 &
    0.055 & 0.28 & 0.0 & 0.33 & -0.91 & 0.92  \\
    \cline{2-5}\cline{7-9} 
    \raisebox{3mm}[0pt]{\ $\sim$1942 clusters} \rule[-3mm]{0mm}{8mm} &
    20\% & 0.65 & 0.060 & 0.30 & & 0.35 & -0.86 & 0.93 \\
    \hline\hline
% Model B (no neutrinos, w = -0.8)
    \rule[-3mm]{0mm}{8mm} & 0\% & 0.64 $^{+ 0.03}_{-0.02}$ & 0.060 $^{+0.004}_{-0.004}$& 0.30$^{+0.02}_{-0.02} $& & 0.36$^{+0.02}_{-0.03} $&
    -0.80$^{+0.06}_{-0.09}$ &\ 0.88$^{+0.03}_{-0.02} $\\ \cline{2-5}\cline{7-9}
    \raisebox{3mm}[0pt]{Model B} \rule[-3mm]{0mm}{8mm} & 10\% & 0.63
    & 0.062 & 0.32 & 0.0 & 0.38 & -0.77 &\ 0.89 \\
    \cline{2-5}\cline{7-9}
    \raisebox{3mm}[0pt]{\ $\sim$ 2248 clusters} \rule[-3mm]{0mm}{8mm} &
    20\% &0.62 & 0.062& 0.32 & & 0.38 & -0.77& 0.92 \\
    \hline\hline
% Model C (with neutrinos, w = -1.0)
   \rule[-3mm]{0mm}{8mm} & 0\% & 0.65 $^{+ 0.01}_{-0.01}$
    & 0.056  $^{+ 0.001}_{-0.002}$& 0.32  $^{+ 0.01}_{-0.01}$& 0.60  $^{+ 0.07}_{-0.13}$& 0.38 $^{+ 0.02}_{-0.01}$ &  & 0.84 $^{+ 0.01}_{-0.02}$ \\ \cline{2-7}\cline{9-9}
      \raisebox{3mm}[0pt]{Model C}    \rule[-3mm]{0mm}{8mm} & 
    10\% & 0.65 & 0.056 & 0.33& 0.50 &0.39 &
  -1.00 & 0.90 \\ \cline{2-7}\cline{9-9}
    \raisebox{3mm}[0pt]{$\sim$ 1601 clusters} \rule[-3mm]{0mm}{8mm} &
       20\% &0.64 & 0.056& 0.33 & 0.50 & 0.40 & & 0.92 \\
    \hline
  \end{tabular}
\caption[Behavior of cosmological parameters with systematic
  error]{Four models showing the change in the best fit cosmological parameters
  for 0\%, 10\%, and 20\% systematic errors in the mass.  In all
  cases, the survey area is $\delta\Omega = 200$ square degrees with a
  limiting mass of $M_{lim} = 2\times 10^{14} M_\odot$ and a single
  mass bin at each redshift slice.
  Models A and B fix the neutrino mass at zero, while model C allows
  the neutrino mass to vary while fixing $w$.  The $1\sigma$
  uncertainty in the fiducial parameters for a Planck-like CMB
  temperature spectrum plus cluster constraints are given to compare
  the against the shifts in parameters from the systematic error. The
  $\sigma_8$ values for all models lie
  within the $2\sigma$ error region obtained from WMAP \cite{wmapy1}.} 
   \label{param_table}
\end{center}
\end{table*}

\begin{figure*}
\begin{center}
  \includegraphics[width=6in]{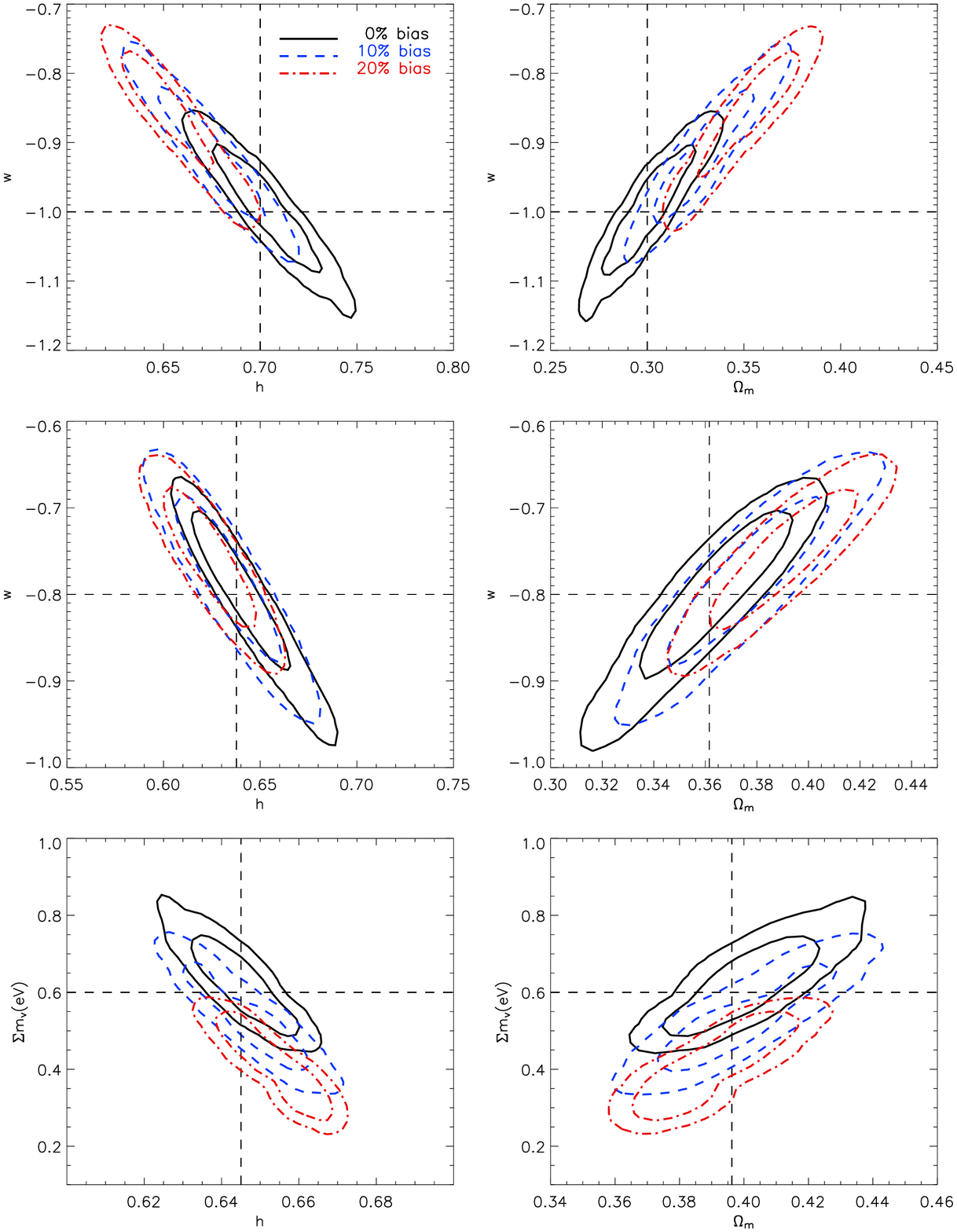}
\end{center}
  \caption[Error contours for 0\% and 10\% systematic error]{Error
  contours for dark energy models, A (top row) and B (middle), and
  massive neutrino model, C (bottom) for 0\%, 10\% and 20\% systematic error
  in cluster mass.} 
  \label{contours}
\end{figure*}

Broadening the parameter space, the uncorrected bias would imply
a larger number of clusters at all masses, as might be
created by an upwards shift in the dark energy equation of state, or
a reduction of neutrino density. In addition to the
$\Lambda$CDM scenario, Table~\ref{param_table} shows parameter fits
for systematic mass
errors for three different models, two in which
the neutrino mass is fixed at zero while $w$ is allowed to vary, and
one in which $\sum m_\nu$ is a parameter while $w = -1.0$ is
fixed.  \fig{contours} shows the error
contours for the dark energy models A and B, and the massive neutrino
scenario, model C, showing the shifts in the peak of the likelihood
distribution as the amount of systematic error in the limiting mass
increases is consistent with an attempt to lessen the suppression in
the growth of structure produced by dark energy and massive neutrinos.

It is clear that the systematic
misestimation of cluster masses can have a significant effect on the
inferred cosmological parameters:  a 10\% shift in mass
can yield parameter shifts on the order of $1\sigma$ (for the
fiducial 200 square degree cluster survey in combination with a
Planck-like CMB experiment). However, the shift in parameters produced
by the bias is highly sensitive to the total number of clusters being
fit. For example, a fiducial neutrino model with $\sigma_{8}$=0.78,
gives only 476 clusters in the survey. In this case, a 10\% mass bias
has an insignificant effect on the best-fit parameters, and the bias
only shows up in the excess $\chi^{2}$ of the fit.

\begin{figure*}
  \includegraphics[height=6in]{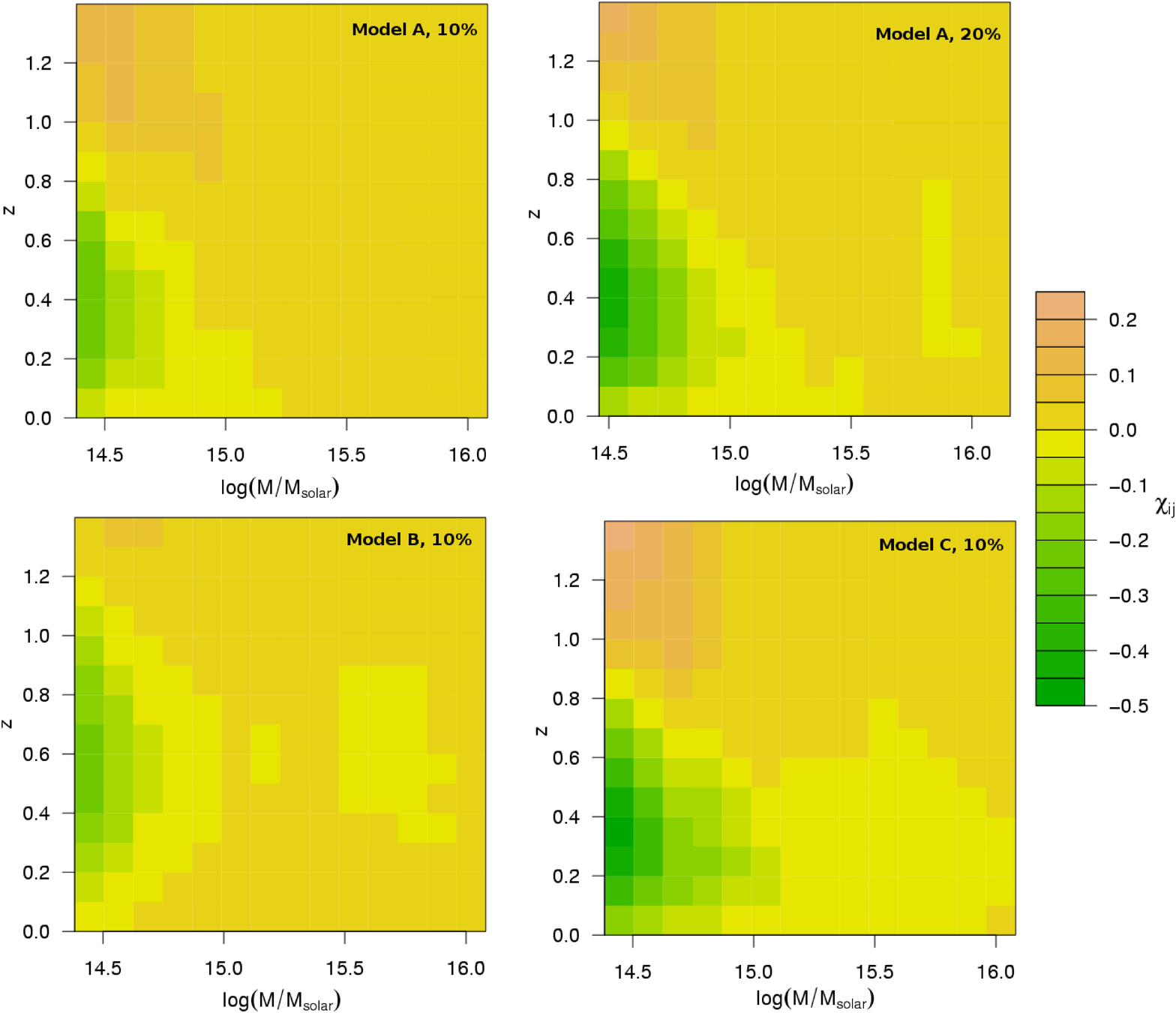}
  \caption[The distribution of redshift and mass bin-wise residuals
  for 10\% systematic error]{The distribution of redshift and mass
  bin-wise residuals $\chi_{ij}$ Models A (top row, 10\% and 20\% mass
  shift), B and C (bottom row, 10\% mass shift).  Note that the distribution of
  positive and negative residuals indicates the change in the shape of
  the mass function due to the change in cosmological parameters. If
  the bin-wise residuals are sufficiently large, this pattern may be a
  means to detect the presence of systematic error in mass estimates.}
  \label{residuals}
\end{figure*}

The goodness of fit of the best-fit model to the mass-biased cluster numbers is
quantified by calculating the residuals
\bea
\chi_{ij} = \frac{ N^\mathrm{fit}_{ij} - N^\mathrm{data}_{ij}}
{\sqrt{N^\mathrm{data}_{ij}}}
\eea
where $N_{ij}$ is the number of clusters in a given mass and redshift
bin, so that $\chi^2 = \sum_{ij} \left( \chi_{ij} \right)^2$.  
\fig{residuals} shows the residuals plotted in mass and redshift bins
for the two models.

One key consideration for the impact of systematics on the
cosmological utility of a given SZ survey is whether the systematic
shift in the measured bin counts is larger than the Poisson error for
the number of clusters in that bin. If so, then the systematic
distortion of the bin counts is detectable and can be measured; if the
systematic shift per bin is smaller than the Poisson error, then it is
not possible to diagnose the systematic error on a bin-by-bin
basis. The bin residual $\chi_{ij}$ is normalized to the Poisson error
in that bin, so if $\chi_{ij} = 1$ in a given bin, the systematic shift
in cluster counts is the same size as the $1\sigma$ statistical
error. The statistical errors scale trivially with the square root of
the survey area $(\delta\Omega)^{1/2}$. In the models we have studied,
\bea
\frac{|\chi_{ij}|}{\sqrt{\delta\Omega}} \lesssim \left\{
\begin{array}{ll} 
%  0.012 \quad & \mathrm{Model\ \Lambda CDM}\\ 
  0.017 \quad & \mathrm{Model\ A} \\ 
  0.015 \quad & \mathrm{Model\ B} \\ 
  0.032 \quad & \mathrm{Model\ C} 
\end{array} \right.
\label{pixel_resid}
\eea
To attain $\chi_{ij}=1$ in the bins with the largest count distortions
requires a survey on the order of 3500 square degrees for model A,
4300 square degrees for model B, and 980 square degrees for model
C. While Planck will cover the
entire sky with perhaps 35000 square degrees usable for cosmology, it
has a higher cluster mass detection threshold ($\simeq
5 \times 10^{14}\, M_\odot$) due to its relatively large beam
\cite{white03}; each Planck cluster bin will have fewer 
clusters, and the values in \eq{pixel_resid} are significantly smaller. 

Figure~\ref{residuals} displays the normalized residuals $\chi_{ij}$
per bin. Not surprisingly, the largest values are at the lower end of
the cluster mass range, where the bins have the largest
populations. This simply reflects the fact that the largest mass
clusters represent the peaks in the initial mass distribution, and
their population is more sensitive to small changes in cosmology. Note
that in all cases considered, the systematic discrepancies are not
randomly distributed throughout the bins, but rather have a 
coherent structure. The condition $\chi_{ij}=1$ should be viewed as a
rough estimate of the overall size of the distortions. By modeling
particular coherent patterns of discrepancy over the mass and redshift
bins, it may be possible to diagnose particular systematic distortions
even if every individual bin has a systematic shift smaller than a
1$\sigma$ statistical error for that bin.

%%%%%%%%%%%%%%%%%%%%%%%%%%%%%%%%%%%%%%%%%%
\section{Discussion}
\label{clust_discussion}
%%%%%%%%%%%%%%%%%%%%%%%%%%%%%%%%%%%%%%%%%%

The compilation of large galaxy cluster catalogs selected
via \SZ distortion of the microwave background
will be a reality within the next few years, and these
data sets will open a new realm of cosmological inquiry. 
The importance and potential impact of these measurements
is widely recognized. A recent report by the
Interagency Working Group on the Physics of the Universe,
based on the National Research Council's 2002 Turner Commission
Report, stated that a
highest priority investigation should be a coordinated effort to use 
the number of galaxy clusters observed in SZ surveys and X-ray 
observations as a dark energy probe \cite{phys_of_univ}. 

The goal of this paper is to provide a realistic assessment of
how well galaxy cluster physics will need to be understood for
the upcoming cluster surveys to realize their cosmological
potential. The cluster properties which are most easily connected
to predictions from cosmological simulations are redshift, mass,
and peculiar velocity. Redshift is, with sufficient optical telescope
resources, measurable to high accuracy. We have therefore not
considered systematic redshift errors here, although for surveys
which rely on photometric redshifts rather than spectroscopic ones
systematic redshift errors may not be negligible. The cluster
mass distribution is most often considered as the basic
relation which future SZ surveys will measure, and here we have
focused on systematic mass errors. The cluster mass is not directly
measurable via  the SZ distortion on an individual cluster basis, and
the relation between SZ flux and mass must be assumed, extracted from
simulations, or measured in some other way. Two general classes of
techniques are currently under study: ``self-calibration'' of the
cluster mass-flux relation directly from the SZ survey
\cite{majumdar2003,lima_hu2005}, and use of other cluster observables
like X-ray emission or weak lensing shear.

It is quite reasonable to expect that a combination of data sets
and analysis techniques will lead to reasonable cluster mass estimators. 
How good will these estimators need to be? Here we have presented a 
model calculation showing that systematic biases in cluster mass
estimates at the 10\% level are enough to shift cosmological
parameter estimates by more than the statistical $1\sigma$
error bars for some parameters, particularly the dark energy
equation of state $w$. This assumes that cosmological parameters
will be constrained using a Planck-like measurement of the
primary microwave background fluctuations. One might hope that
the distribution of cluster masses and redshifts would be sufficiently
altered by systematic misestimates of cluster masses that the systematic
error would be detectable in the cluster distribution itself;  
we show here that this is likely to be only marginally possible with
upcoming cluster SZ surveys.

Our conclusions from this study are cautiously optimistic. It is
clear that upcoming SZ cluster surveys will be in the regime where
systematic errors due to cluster astrophysics will be important for
interpreting the results. On the other hand, with a variety of
potential observations and techniques for diagnosing and accounting
for systematic errors, we can plausibly expect to reduce the
impact of systematic errors on cosmological conclusions to the
level of statistical ones. This will by no means be simple; 
cluster mass estimates for a large sample with no more than 10\%
bias is hard to do. (Note that the bias in the cluster mass
measurements, and not the size of the scatter in the measurements, is
the relevant error to consider.) SZ cluster catalogs must be
conceptualized as the basis for a range of other complementary
measurements and calculations which, taken as a whole, can
contribute significant cosmological constraints.

The challenges of controlling systematic errors in cluster mass
estimates also prompts consideration of alternate possibilities
for extracting cosmology from SZ catalogs. The underlying difficulty
with cluster masses is that they are only indirectly probed by
SZ measurements, but the directly measured SZ flux is not easily
related to cosmological properties (and the clusters will likely not
even be precisely flux-selected \cite{melin2005}). Furthermore, most
astrophysical processes in clusters shift the SZ signal towards
larger fluxes, so mass estimates from the SZ signal are generally biased
high, and a careful accounting for all contributing effects must be
undertaken. The number of clusters as a function  of redshift or mass
is very sensitive to this unavoidable difficulty.  Other cluster
observables are potentially less sensitive to mass biases in an SZ
cluster catalog. The distribution of clusters on the sky and in
redshift is one obvious possibility
\cite{wkhm2004,majumdar2003,mei03,mei04}. While selection biases in
large-scale structure surveys have been studied extensively in the
context of galaxy catalogs, relatively little analysis has been done
on selection biases in corresponding SZ-selected cluster catalogs. 

Another alternative may be to use the kinematic SZ effect to construct
cluster peculiar velocity catalogs \cite{dore03}.  The kinematic SZ
effect is smaller amplitude than the thermal SZ signal, and its
frequency dependence is nearly the same as the blackbody primary
microwave background fluctuations, so its detection requires higher
sensitivity measurements and sophisticated techniques for separating
the kinematic SZ signal from other signals and noise sources. But its
advantage is that few systematic errors are correlated strongly with
the kinematic SZ signal, and unbiased estimates of cluster peculiar
velocities are possible in principle \cite{nagai03,sehgal05}.  This is
a promising alternate route to cosmological constraints from SZ
cluster catalogs which is less susceptible to systematic errors
intrinsic to cluster properties \cite{DeDeo05,Hernandez05}.

The goal of this paper is to shift the focus of the discussion about
SZ surveys from their abundant potential to provide interesting
constraints on cosmology, which has been well demonstrated, to
the level at which systematic errors must be controlled. Some systematic
errors are unavoidable, due to intrinsic astrophysical uncertainties
in galaxy clusters, while others will result from practical limitations
on given experiments and on algorithms for separating different signal
components given a limited number of frequency channels and limited
angular resolution. Here we advocate, in addition to detailed study of
these individual effects, examining the impact of all of these using
an effective model of systematic errors in the ultimate physical
quantities used in constructing cosmological tests, namely cluster
masses and redshifts. We have considered a simple proportional shift in
inferred cluster mass relative to the actual cluster mass; clearly
more complex models may be useful. We have also looked only at a few
cosmological models, due to the computational difficulty of surveying
wide sets of models each with its own set of Markov Chain Monte Carlo
calculations. One highly useful direction for future work is
constructing much faster approximations to the matter power spectrum
(especially in the massive neutrino case), which could greatly
increase the Markov Chain efficiency. Such approximations have already
been proven for  the microwave background primary power spectrum
\cite{kmj,cmbwarp,kaplinghat02}, which is more complicated than the
matter power spectrum. With such tools in hand, it will be possible to
perform far more exhaustive calculations of systematic effects than
those presented here, including a wider range of underlying
cosmological models, different models of systematic effects, and
combinations of other sources of cosmological data.  This kind of
extensive analysis is, in our opinion, essential for planning
observational programs complementary to the SZ surveys currently
undertaken, programs which will be demanding in time and resources yet
which hold the key to maximizing the return on our investment in SZ
observations.

\begin{acknowledgments}

MRF thanks Jack Hughes and Licia Verde for their many helpful comments on
the chapter of his thesis that parallels this paper, and Antony Lewis
for help in using code from CAMB and CosmoMC.  The authors thank Raul
Jimenez, Lyman Page and an anonymous referee for useful comments.  
This work has been partly supported by the NSF award 0408698 for the
ACT project. The work of RB was supported by funds from Cornell
University. This research used computational facilities supported by
NSF grant AST-0216105.

\end{acknowledgments}

\bibliography{cluster}

\end{document}